\documentclass[psfig,usenatbib]{mn2e}
\input psfig.sty
\input{epsf}

\def\lsim{\mathrel{\hbox{\rlap{\hbox{\lower4pt\hbox{$\sim$}}}\hbox{$<$}}}}
\def\gsim{\mathrel{\hbox{\rlap{\hbox{\lower4pt\hbox{$\sim$}}}\hbox{$>$}}}}

\newcommand{\aap}    {A\&A}
\newcommand{\apjs}   {ApJS}
\newcommand{\apj}    {ApJ}
\newcommand{\apjl}   {ApJL}

\newcommand{\mnras}  {MNRAS}

\title[Large scale anisotropies on halo infall]{Large scale anisotropies on halo infall}
\author[L. Ceccarelli et al.]{
\parbox[t]{\textwidth}
{ Laura Ceccarelli$^{1,2}$\thanks{E-mail:ceccarelli.laura@gmail.com},
  Dante J. Paz$^{1,2}$,
  Nelson Padilla$^3$,
  Diego G. Lambas$^{1,2}$
}
\vspace*{6pt}\\
$^1$ Instituto de Astronom\'\i a Te\'orica y Experimental, UNC-CONICET, C\'ordoba Argentina. \\
$^2$ Observatorio Astron\'omico de C\'ordoba, UNC, Argentina. \\
$^3$ Departamento de Astronom\'\i a y Astrof\'\i sica, Pontificia
     Universidad Cat\'olica de Chile, Santiago, Chile.\\
}

\begin{document}

\date{\today}

\maketitle

\begin{abstract}

We perform a statistical analysis of the peculiar velocity field around dark
matter haloes in numerical simulations. We examine different properties of the
infall of material onto haloes and its relation to central halo shapes and the
shape of the large scale surrounding regions (LSSR).

We find that the amplitude of the infall velocity field along the halo shape
minor axis is larger than that along the major axis. This is consistent for
general triaxial haloes, and for both prolate and oblate systems. We also
report a strong anisotropy of the velocity field along the principal axes of
the LSSR.

The infall velocity field around dark matter haloes reaches a maximum value
along the direction of the minor axis of the LSSR, whereas along the direction
of its major axis, it exhibits the smallest velocities. We also analyse the
dependence of the matter velocity field on the local environment. The amplitude
of the infall velocity at high local density regions is larger than at low
local density regions.  The velocity field tends to be more laminar along the
direction towards the minor axis of the LSSR, where the mean ratio between flow
velocity and velocity dispersion is of order unity and nearly constant up to
scales of 15 Mpc h$^{-1}$.

We also detect anisotropies in the outflowing component of the velocity field,
showing a maximum amplitude along the surrounding LSSR major axis. 

\end{abstract}

\begin{keywords}
large scale structures: overdensities: infall , statistical, haloes 
\end{keywords}

\section{Introduction}

Given the current paradigm for structure formation in the Universe, mass is
accreted onto haloes from the network of filament walls and voids.  Therefore,
a strong correlation between halo properties and large scale dynamics is
expected in this scenario.  Even though statistical isotropy at very large
scales is a fundamental assumption in cosmology widely verified in galaxy
surveys, on smaller scales of a few Mpc, the structure of the mass distribution
departs significantly from spherical symmetry with the usual presence of walls,
filaments and voids. This anisotropy in the matter distribution has been
extensively studied, as it is related to the search for large scale structures
in the environment close to galaxies. For example, both observational studies
and numerical analysis \citep[e.g.][]{1994West, 2002Plionis, 2003Kitzbichler,
2002Faltenbacher} reported that galaxies tend to be aligned with their
neighbours supporting the vision of anisotropic mergers along filamentary
structures. When considering preferential directions along the large scale
structures, it is natural to underpin long filamentary structures connecting
large clusters.  Regarding the evolution of structure in simulations, the flow
of particles and haloes within these filaments is apparent and can originate
preferential directions in the velocity field
\citep[e.g.][]{2010Gonzalez,2006Pivato}. 

On the other hand, the anisotropic distribution of structure around haloes,
induces the emergence of vortical motions on the halo velocity field, through
tidal interactions. Previous works have shown that the distributions of spin
vectors are not random. For example, the spin of haloes in simulations tend to
point in the direction orthogonal to the filaments  \citep[see][ and references
therein]{PaperL}. Combined with the results suggesting that halo spins are more
sensitive to recent infall \citep{1993vanHaarlem_Fields}, these alignment
properties fit well with accretion along preferred directions. In addition,
several authors have shown that halo shapes are aligned with the surrounding
large scale matter distribution, with increasing ellipticity as the mass
increases \citep[see for instance][]{PaperShapes}. Therefore halo shapes are
likely to depend on the accretion process driven by halo mergers. Following
this line, it is expected a strong correlation between the velocity field in
the neighborhood of haloes and the directions defined by the halo shape axes.

Most of these previous studies focused on the fact that alignments and
preferential directions are consequences of the formation process of haloes.
However, the effects of such preferential directions on the peculiar velocity
field at large scales have not been so extensively addressed. In a hierarchical
clustering scenario these peculiar velocities are generated by inhomogeneities
in the distribution of matter in the universe. The nature of this velocity
field depends on the local density. High density regions show random motions
typical of virialised objects whereas low density environments, on the other
hand, are more likely to show streaming motions: objects falling towards larger
potential wells constantly increasing the amplitude of their clustering
strength \citep{1997Diaferio}.

In this work we use a numerical simulation of the standard scenario of Cold
Dark-Matter with a cosmological constant ($\Lambda$CDM) to explore different
features of the peculiar velocity field and its relation to haloes and their
large-scale surrounding region (referred to as LSSR hereafter).  In Section 2
we describe the analysis of the simulation, Section 3 deals with the
anisotropies of the velocity field in relation to halo and LSSR shape and
orientation.  We also consider in this section  the infall of particles and
haloes in different local density regions.  In Section 4 we explore the
turbulence of the flow along different directions and finally in Section 5 we
analyse the outflow of particles from haloes and its relation to the LSSR.  We
present our conclusions in Section 6.

\section{Data }

Throughout this work we use a collisionless numerical simulation of
$512^3$ particles covering a periodic volume of $500^3$ $($h$^{-1} {\rm
Mpc})^3$. We assume a spatially flat low-density Universe, with a matter
density $\Omega_{\rm m}=1-\Omega_{\Lambda}=0.28$, Hubble constant
$H_{\circ}=74$ km s$^{-1}$ Mpc$^{-1}$, and normalisation parameter
$\sigma_{8}=0.8$. With these parameters the resulting particle resolution is
$m_{\rm p} =  7.2\times 10^{10}\,$M$_{\odot}$h$^{-1}$. The initial
conditions at redshift $\sim 50$ were generated with the {\small GRAFIC2}
package \citep{2001Bertsch}, which also computes the transfer function as
described in \citet{1995Ma}. The simulation run was performed using the second
version of the {\small GADGET} code developed by \citet{Gadget2}, with a
gravitational softening of $0.03 $h$^{-1} {\rm Mpc}$ chosen following
\citet{PowerC}. 

The identification of particle clumps to be associated to dark matter haloes
was carried out by means of a standard friends-of-friends algorithm with a
percolation length given by $l=0.17$ $\bar{\nu}^{-1/3}$, where $\bar{\nu}$ is
the mean number density of dark matter particles. We identified approximately
$470,000$ dark-matter haloes with at least $10$ particle members, resulting in
a minimum halo mass of $M_{\rm min}=7.2\ 10^{11}$M$_{\odot}$h$^{-1}$. It
should be noticed that this algorithm does not resolve substructures within the
dark-matter haloes.

From this point on, we will centre most of our analysis on samples of
central haloes with different masses.  We draw five halo subsamples with
equal number of haloes of different mass starting at a minimum of $50$
particles or $3.5 \times 10^{12}$M$_{\odot} $h$^{-1}$ (the lowest mass sample
considered includes haloes of up to $64$ particles each).  The subsample with
the most massive haloes starts at $260$ particle members corresponding to $1.9 \times 10^{13} $M$_{\odot}$h$^{-1}$.

Throughout this work, we estimate statistical errors by means of the jacknife
technique, which has been shown to provide equivalent results to error
estimates obtained from the variance of a large number of independent
simulations \citep[see for instance,][]{Croton2004,2003PadillaBaugh}.  These
results indicate that jacknife errors provide a reasonable estimate of
statistical uncertainties and cosmic variance.

\section{Velocity field anisotropies}

In this section, we analyse anisotropies of the radial velocity field relative
to the centre halo shape axes. We also analyse the dependence of this velocity
field with respect to the surrounding mass distribution at small and large
scales.

\subsection{Dependence on the halo shape}

Several authors \citep[][and references therein]{1992Warren, 1998Thomas,
2005Hopkins, 2006Allgood_Shape, 2005Kasun_Shape, LauShapes, PaperShapes} have
analysed the properties of halo shapes using the best-fitting ellipsoid to the
spatial distribution of the halo material. Following this standard method, we
calculate the shape tensor for each dark-matter halo using the positions of its
particle members. This can be written as a symmetric matrix,
\begin{equation}
I_{ij}= (1/N_h)\sum_{\alpha=1}^{N_h} X_{\alpha i} X_{\alpha j},
\label{eq:it}
\label{eq:shape} \end{equation}
where $X_{\alpha i}$ is the $i^{th}$ component of the displacement vector of a
particle $\alpha$ relative to the centre of mass, and $N_{h}$ is the number of
particles in the halo. The matrix eigenvalues correspond to the square of the
axis ($a$, $b$, $c$ were $a>b>c$) of the characteristic ellipsoid that best
describes the spatial distribution of the halo members. The principal axes of
this ellipsoid are represented by the eigenvectors $\hat{a}$, $\hat{b}$ and
$\hat{c}$\footnote{A fixed $b/a=1$ with an arbitrary value of $c/b$,
corresponds to perfect oblate ellipsoids. On the other hand, systems with fixed
$c/b=1$ are perfect prolate ellipsoids}. Haloes with $b/a<c/b$ are associated
to general triaxial ellipsoids with prolate shape, while the opposite case,
$b/a>c/b$, corresponds to predominantly oblate systems.  Numerical results
based on this shape analysis have shown that haloes exhibit aspherical shapes,
with a slight preference toward prolate forms.  Moreover, it has been shown
that their orientations are related to their surrounding structures
\citep{2005Colberg_Shape, 2005Kasun_Shape, 2006Basilakos_Shape,
2006Allgood_Shape, 2006Altay_Shape, 2007Aragon_Spin, 2007Brunino_Shape,
2007Bett_Shape, 2009Zhang_Spin}. Several authors have found that dark matter
haloes tend to be more prolate and aspherical when larger halo masses are
considered \citep[][and references therein]{2005Kasun_Shape,PaperShapes}. This
trend could be easily understood as the result of an accretion process driven
by halo mergers. Therefore, we have a physical motivation to expect a
connection between the halo surrounding velocity field and the particular
directions defined by the halo shape axes.

\begin{figure}
    \epsfxsize=8cm
    \centerline{\epsffile{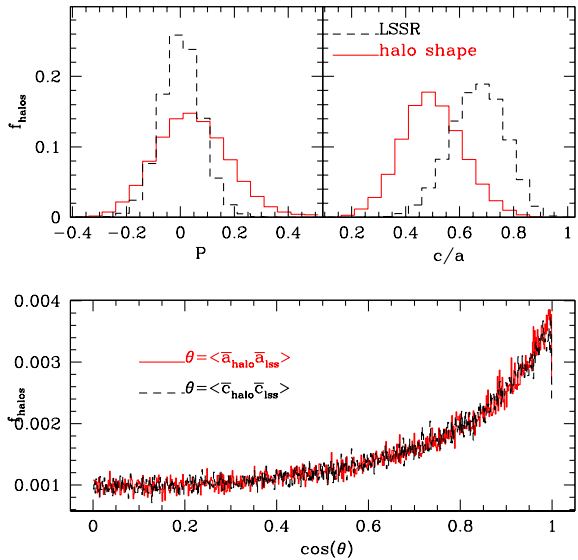}}
    \vspace*{-.2cm}
    \caption
    { 
	Upper panel: histograms showing the shape (left) and eccentricity
	distribution (right) of haloes (red) and the LSSR
	(black).  Lower panel: normalised distribution of the angle between
	haloes and LSSR minor axes (black). The corresponding distribution for
	major axes is shown in red lines.  
    }
    \label{fig:fhist}
\end{figure}

The resulting halo ellipticities, which are defined as the ratio $c/a$, are
shown in red lines in the right upper panel  of Figure \ref{fig:fhist}.
Following \citet{PaperShapes} we separate prolate and oblate ellipsoids through
a triaxiality parameter ``$P$'' defined as:
\begin{equation}
P=log\left( (c/b)/(b/a)\right),
\label{eq:p1} 
\end{equation}
where systems characterised by $P >0$ ($P<0$) correspond to halo with prolate
(oblate) shapes. In the left upper panel of Figure \ref{fig:fhist} we show the
distribution of triaxiality parameter values (red line) derived for haloes.  As
can be noticed in this figure haloes show a trend towards prolate shapes, in
agreement with previous results \citep[see for instance][]{PaperShapes}

In order to characterise velocity field anisotropies around haloes, we
calculate the mean radial velocity of particles taking into account their
positions relative to the dark matter halo major axes. We compute radial
velocities relative to the halo centre, with positive velocities indicating
mean infall flux and negative values, outflow velocities. Along this subsection
we analyse elongated haloes, characterised by small ellipticity parameters
($e=c/a < 0.5$). From this subsample we select either prolate or oblate haloes,
corresponding to triaxiality parameters $P > 0.1$ or $P < -0.1$. From this
point on we will simply refer to these two halo subsamples as prolate and
oblate haloes.

\begin{figure*}
    \epsfxsize=12.0cm
    \centerline{\epsffile{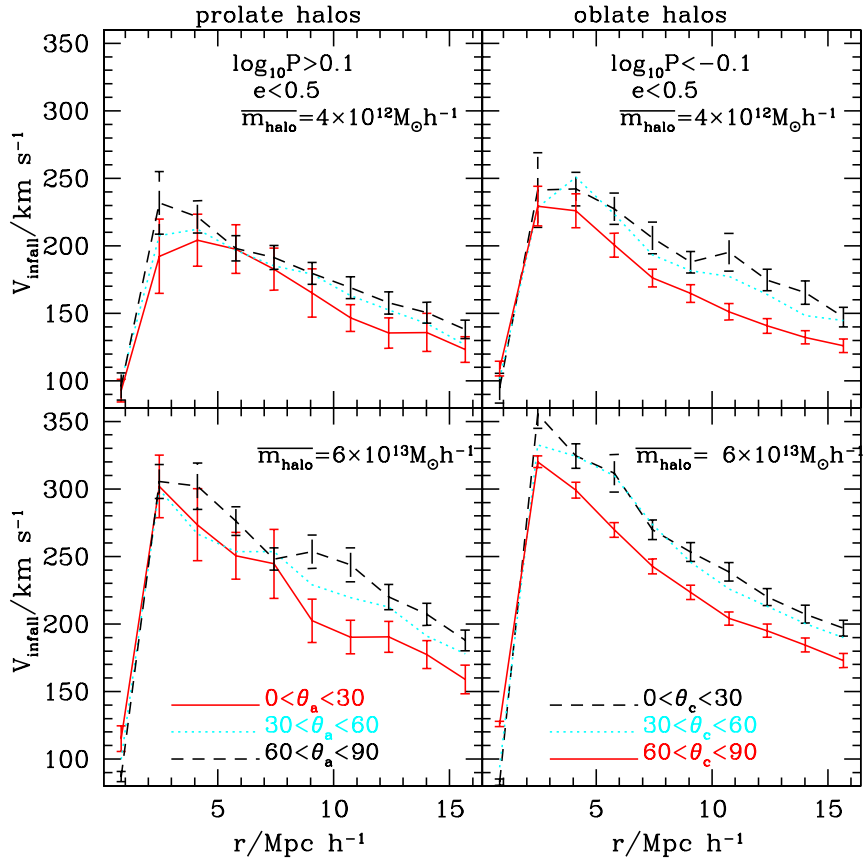}}
    \caption
    {
      Mean infall velocity as a function of distance to the central halo for
      different shape of halo.  Upper panels correspond to high mass haloes
      and lower panels to low masses as is indicated in the figure.  Left panels
      show results for prolate haloes and right panels for oblate haloes.
      Different colours indicate infall directions at different angles from 
      the halo principal axes $a$ and $c$.
      Red lines in the left panels indicate $\theta_{a}<\pi/$6, cyan lines
      $\pi/$6$<\theta_{a}<\pi/$3 and black lines $\pi/$3$<\theta_{a}<\pi/$2.
      Black lines in right panels indicate $\theta_{c}<\pi/$6, cyan lines
      $\pi/$6$<\theta_{c}<\pi/$3 and red lines $\pi/$3$<\theta_{c}<\pi/$2.
    }
    \label{fig:f1}
\end{figure*}

For prolate haloes, we define $\theta_{a}$ as the angle subtended by the
halo-particle relative position and the major axis. We calculate the average
radial velocity as a function of halo-particle separation in three cases; i)
for particles close to the major axis ($\theta_{a}<\pi/6$), ii) for particles
roughly perpendicular to the major axis ($\pi/3<\theta_{a}<\pi/2$) and iii) for
particles at intermediate positions ($\pi/6<\theta_{a}<\pi/3$). The resulting
velocity curves are shown in the left panels in Figure \ref{fig:f1}, where the
upper panel corresponds to low mass haloes ($\bar{m}_{halo}=4\times
10^{12}$M$_{\odot}$h$^{-1}$; $7.2\times 10^{11}$M$_{\odot}$h$^{-1}<m_{halo}<
4.5\times 10^{12}$M$_{\odot}$h$^{-1}$) and the lower panel corresponds to high
mass haloes ($\bar{m}_{halo}=6 \times 10^{13}$M$_{\odot}$h$^{-1}$; $1.9\times
10^{13}$M$_{\odot}$h$^{-1}<m_{halo}$).  Red lines in the left panels Figure
\ref{fig:f1} show the velocity field traced by particles along the major axis ;
cyan lines correspond to particles in intermediate positions and black lines
correspond to velocities of particles with positions perpendicular to the major
axis.  Similar results were obtained for the intermediate mass bins.

In a similar fashion, we derive the velocity field for particles along the
minor axis and in the perpendicular direction for oblate haloes.  We define
$\theta_{c}$ as the angle subtended by the halo-particle relative position and
the minor axis direction. The mean velocities obtained are shown in the right
panels of Figure \ref{fig:f1} where black lines show the velocity field traced
by particles along the minor axis ($\theta_{c}<\pi/$6), red lines correspond to
velocities of particles with positions perpendicular to the minor axis
($\pi/$3$<\theta_{c}<\pi/$2), and cyan lines correspond to particles at
intermediate positions ($\pi/$6$<\theta_{c}<\pi/$3). We use the same halo mass
ranges in both left and right panels of Figure \ref{fig:f1}. 

As can be seen in Figure \ref{fig:f1}, velocities are positive over the entire
scale range, indicating that the surrounding material is infalling onto haloes
regardless its mass or direction. It is worth mentioning that the larger infall
velocities are obtained for the more massive haloes, in agreement with previous
results \citep{Lauinfall} and theoretical predictions \citep[][and references
therein]{2006Pivato,Crofto}.
 
Regarding the analysis of prolate haloes, it can be seen in the left panels of
Figure \ref{fig:f1}, that the mass infall amplitude is sightly smaller in the
direction along the major axis (red lines) than the corresponding infall
obtained over the perpendicular plane to major axis (black lines).  For oblate
haloes (right panel on Figure \ref{fig:f1}), the largest infall velocities are
obtained at directions parallel to the minor halo axis (black line), which
exceed by approximately $40$ km/s the corresponding velocities obtained along
the perpendicular direction (red line). Both observed anisotropies for prolate
and oblate halo subsamples, can be understood as a tendency of the surrounding
mass to fall faster from minor axis directions, which are the directions set by
the more flattened sides of the halo shape ellipsoid.  This difference is
detected up to distances to the halo centre of the order of a few times the
virial radius (approximately up to distances around $2$ to $3$ Mpc h$^{-1}$).

\subsection{Dependence with the surrounding structure}

We have also analysed the velocity field dependence on directions defined over 
the surrounding large scale mater distribution. For this purpose we estimate
the shape tensor of a spherical large scale region around dark matter haloes,
excluding the inner region containing the central halo. In the same way as for
dark matter haloes, the surrounding large scale distribution can be
characterised by triaxial ellipsoids with positive or negative triaxiality
parameters. It can be said that LSSR
with $P>0$ ($P<0$) correspond to ``filamentary'' (``sheet-like'') matter
distributions. We use this surrounding large-scale structure at
the present time ($z=0$) to characterise the large-scale environment. We expect the
presence of correlations between the LSSR distribution, and the
surrounding velocity field at the present time.  We stress the fact that the
dynamical timescale of this volume is large enough to contain information
on the accretion process of the central halo across the halo evolutionary
stages. Thus, compared to the halo crossing time, the LSSR inertia
principal axes can be taken as quasi stationary. For instance, a typical halo
of $10^{13}\,{\mathrm M}_\odot$ has a velocity dispersion around $300$
km/s, with a crossing time at virial radius of $1.4$ Gyrs, whereas a
particle with typical velocity of $400$ km/s takes $75$ Gyrs to reach a position $10$
Mpc h$^{-1}$ distant.

As described 
above, we characterise the LSSR by means of the shape tensor as defined 
in equation \ref{eq:it}.  For the calculation we use particles 
between $2$ and $15$ Mpc h$^{-1}$ from halo centres, with a minimum separation
threshold so as to avoid the central halo particles. In order to fix the
maximum distance of this spherical shell we have considered several thresholds
and  finally selected the specific value that shows the stronger signal.
Nevertheless we obtain similar results for the different distances considered.
The principal axes are labelled with $\hat{A}$, $\hat{B}$ and
$\hat{C}$, for the major, intermediate and minor axes, respectively. We use
capital letters in order to avoid confusion with the axes $\hat{a}$,
$\hat{b}$ and $\hat{c}$, which are used to characterise the intrinsic halo
shapes.   Given that the LSSR shape tensor represents the second moment of the
matter distribution around a given halo centre, the $\hat{A}$ axis corresponds to 
the direction along the densest direction, whereas the $\hat{C}$
axis corresponds to the lowest density regions.

\begin{figure*}
    \epsfxsize=12.0cm
    \centerline{\epsffile{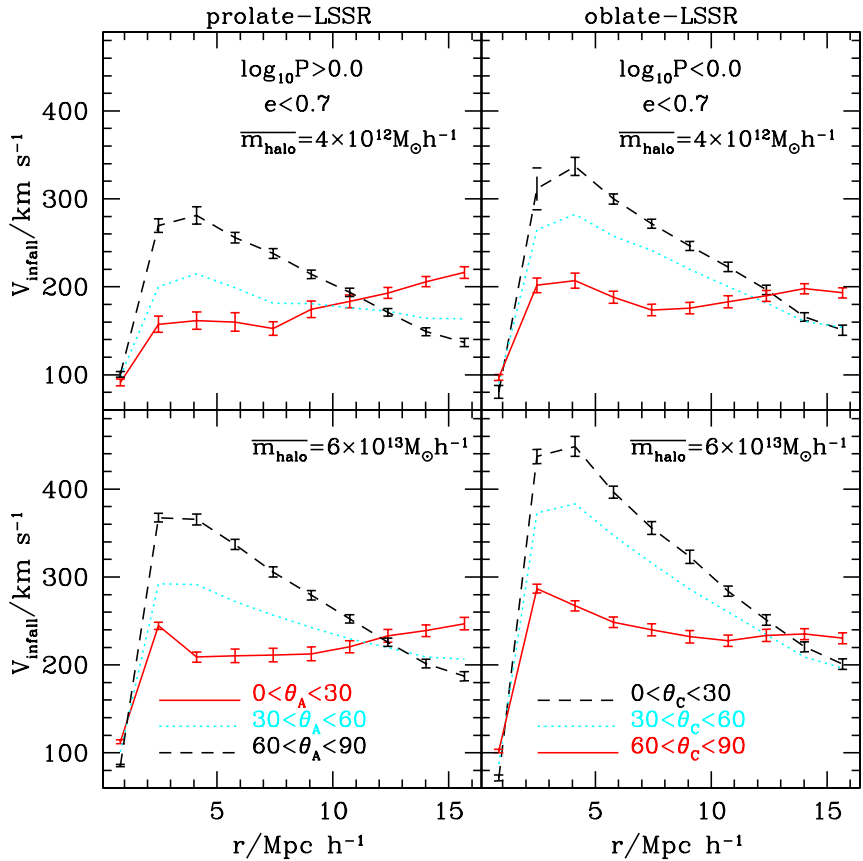}
    }
    \caption
    {
    Mean infall velocity as a function of distance to the central halo for
    different LSSR.  The upper panels correspond to high mass haloes
    and the lower panels to low masses as is indicated in the figure.  Left panels
    show results for prolate-LSSR and right panels for oblate-LSSR.  Different
    colours indicate infall directions at different angles from the LSSR principal axes
    $A$ and $C$.  Red lines
    in the left panels indicate $\theta_{A}<\pi/$6, cyan lines
    $\pi/$6$<\theta_{A}<\pi/$3 and black lines $\pi/$3$<\theta_{A}<\pi/$2.  Black
    lines in the right panels indicate $\theta_{C}<\pi/$6, cyan lines
    $\pi/$6$<\theta_{C}<\pi/$3 and red lines $\pi/$3$<\theta_{C}<\pi/$2.
    }
    \label{fig:f3}
\end{figure*}

As in the previous section (see equation \ref{eq:p1}), we define a triaxiality
parameter for the surrounding distribution using the eigen-values of the LSSR
shape tensor, $P=log((C/B)/(B/A))$.  In the top panels of Figure
\ref{fig:fhist} we show the distribution of $P$ and $C/A$ (dashed black
lines). The cases with positive parameter, $P > 0$, (i.e. $C/B >> B/A$, and
then $C < B << A$) correspond to structures resembling to some extent
ellipsoidal prolate distributions, hereafter we shall refer to such regions
simply as prolate LSSR; filamentary structures are included in this category.
On the other hand, the condition $P < 0$ (i.e. $B/A >> C/B$, so $C << B < A$ )
selects LSSR regions which resemble oblate structures. Hereafter we shall call
such regions simply oblate LSSR.  Structures such as pancakes or walls are
examples of this class. We acknowledge that through this analysis of
triaxiality parameter we cannot obtain a complete topological description of
the LSSR as would be obtained, for example, using a Minkowski functional
analysis. However, given the statistical scope of the current analysis, we
restrict our study to regions with large ellipticities ($C/A < 0.7$), in order
to diminish under-determinations of the corresponding characteristic axes $A$
and $C$ for both prolate and oblate LSSR samples.  In order to assess the
alignment between the LSSR distribution and the halo shapes we compute the
angle $\theta$ between their major axes; the result is shown in red lines in
the lower panel of Figure \ref{fig:fhist}. Similarly the same panel shows the
angles between the minor axis of halos and LSSR (black lines).  As can be seen,
the major and minor axes of the halo shape and the LSSR are highly aligned, in
qualitative agreement with previous results \citep{2005Colberg_Shape,
2005Kasun_Shape, 2006Basilakos_Shape, 2006Allgood_Shape, 2006Altay_Shape,
2007Aragon_Spin, 2007Brunino_Shape, 2007Bett_Shape, 2009Zhang_Spin}. 

Using a similar analysis to that performed in the previous subsection we
compute averages of the radial velocity field as a function of the distance to
the halo centres for different directions. The direction is selected using the
angle ($\theta$) subtended by the halo-to-particle position vector with respect
to the axes $\hat{A}$ or $\hat{C}$, for prolate or oblate LSSR, respectively.
The results are shown in Figure \ref{fig:f3}.  Depending on the triaxiality
parameter of the LSSR region, the mean velocity is computed taking into account
particles with angles $\theta <\pi/6$ with respect to the axis $\hat{A}$ and
$\hat{C}$, red (black) lines in left (right) panels of Figure \ref{fig:f3}.  On
the other hand, we also compute the mean radial velocity field in the
perpendicular directions to these axes requiring $\pi/3 <\theta<\pi/2$, black
(red) lines in left (right) panels of Figure \ref{fig:f3}. Intermediate
positions for both, prolate and oblate LSSR shapes, are obtained in the
interval $\pi/6<\theta<\pi/3$ (cyan lines).

The left and right panels in Figure \ref{fig:f3} correspond to prolate and
oblate LSSR regions, respectively, whereas the upper and lower panels show
results for low and high mass haloes, respectively.  For simplicity we
only show the results corresponding to the lowest and highest mass samples.
As can be seen in the left panels of Figure \ref{fig:f3} (prolate environments)
particles perpendicular to the major axis (black lines) exhibit noticeably
higher velocities than particles parallel to this axis (red lines). A similar
behavior can be seen in the right panels on Figure \ref{fig:f3} (oblate LSSR),
where particles tend to fall faster from directions parallel to the minor axis
$\hat{C}$. Both results are consistent and indicate the preference for a small
velocity infall along the characteristic axis of both, oblate and prolate
surrounding structure. In other words, particles show a faster infall when they
come from low density regions which can be related to the direction of the
smaller shape tensor eigenvalues.  The mean infall velocity reaches a maximum
of approximately $300$ km/s  and exhibits values higher than $400$ km/s for the
more massive haloes. As the distance from halo centre increases, the infall
velocity monotonously decreases. Along the major axis direction, particles
arrive slower with radial velocities that approach constant values or even
increase at greater distances.

\subsection{Local density and the LSSR }

\begin{figure*}
   \begin{picture}(430,250)
      \put(-40,0){\psfig{file=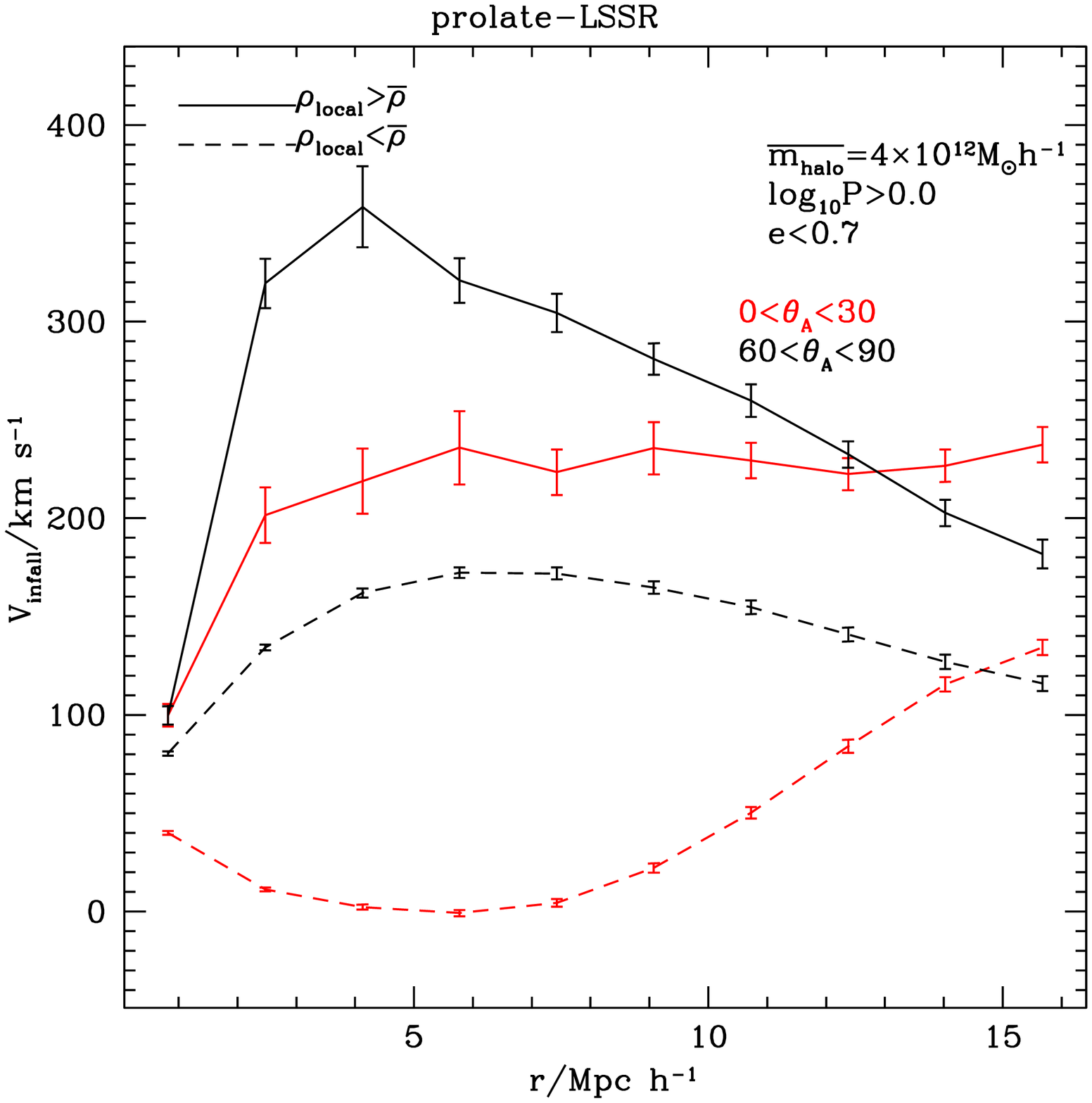,width=9.cm}}
      \put(210,0){\psfig{file=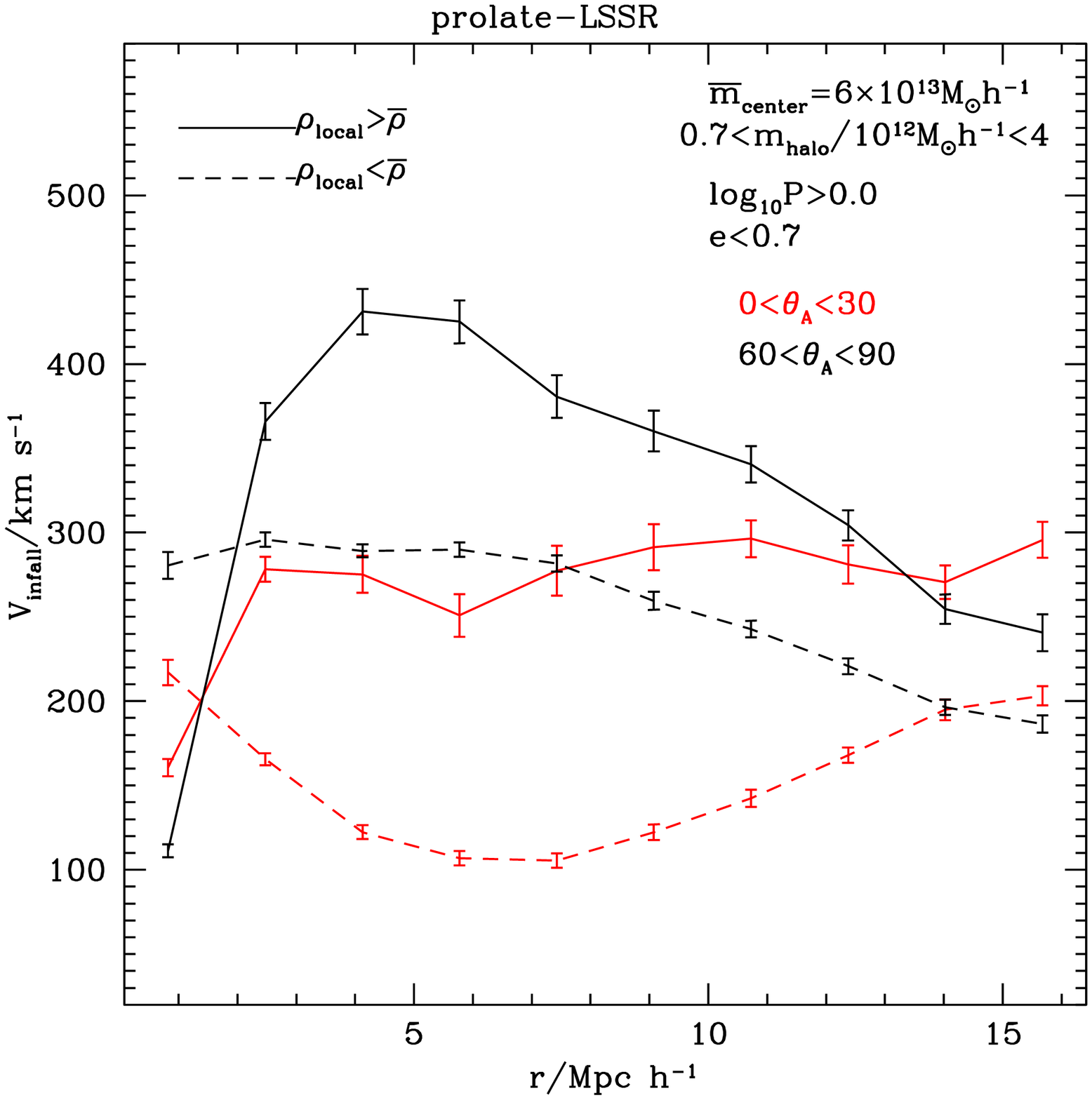,width=9.cm}}
   \end{picture}
   \caption
   { 
       Left panel: Mean infall velocity as a function of distance for particles
       around  low mass haloes ($3.5 \times 10^{12} $M$_{\odot}$h$^{-1} < m_{halo} < 4.5
       \times 10^{12} $M$_{\odot}$h$^{-1}$) and prolate-LSSR ($C/A < 0.7$ and $P > 0.0$) .
       Solid lines correspond to local densities higher the mean local density
       and dashed lines correspond to local densities lower than mean, as is
       indicated in the figure. Red lines indicate $\theta_{A}<\pi/$6, whereas
       black lines indicate $\pi/$3$<\theta_{A}<\pi/$2.  Right panel: Mean
       radial velocity as a function of the distance for    low mass haloes
       ($7.2\times 10^{11}$M$_{\odot}$h$^{-1}<m_{halo}< 4.5\times 10^{12}$M$_{\odot}$h$^{-1}$)
       around high mass haloes ($1.9\times 10^{13}$M$_{\odot}$h$^{-1}<m_{halo}$) and
       prolate-LSSR ($C/A<0.7$ and $P>0.0$).  Lines and colours are as in the 
       left panel.
   }
\label{fig:f5}
\end{figure*}

It is widely accepted that mass follows anisotropic streaming motions toward
clusters, preferentially via the filaments \citep{1999Colberg, 1997Splinter,
1993Haarlem}.  Also, some authors claim that higher density particles exhibit
higher infall velocities and coherence than particles in lower density
environments \citep[see for instance][]{2006Pivato,roberto_nlsn}. This
behaviour seems to be opposite to the correlation observed in the previous
subsection, where we found larger infall velocities along the minor axes for
both, haloes and the surrounding LSSR. Therefore, we now explore the dependence
on the local density field of the mean radial velocities and their anisotropy. 

 We define a local density parameter ($\rho$) for each particle, which
measures the mass density in a sphere of radius equal to the
distance to the 100th nearest neighbour, centred around the particle.  
We also define a mean local density
$\bar{\rho}$ as the average of density parameter logarithms ($Log(\rho)$),
its value corresponds to $5$ particles$/Mpc^3$ for this simulation.

We separate the particles in the simulation according to their associated local
density, into high and low density samples, defined by $\rho > \bar{\rho}$
and $\rho < \bar{\rho}$, respectively.  The left panel of Figure
\ref{fig:f5} shows the mean radial velocity as a function of the distance to
halo centres, for particles in low and high density regions.  We also analyse
the dependence of these results on the LSSR axis directions.  For simplicity we
only show results for prolate-LSSR. 

The velocity curves shown in this figure correspond to low mass central haloes
($3.5 \times 10^{12} $M$_{\odot}$h$^{-1}<m_{halo}<4.5 \times 10^{12}
$M$_{\odot}$h$^{-1}$), solid lines represent results obtained by using particles in
dense local environments ($\rho > \bar{\rho}$), and dashed lines correspond
to particles in underdense local regions ($\rho < \bar{\rho}$).  The
different colours in Figure \ref{fig:f5} correspond to different LSSR
directions; red lines show the mean velocity for particles close to the major
LSSR axis ($\theta_{A}<\pi/$6) whereas black lines show the mean velocity for
particles with $\pi/3<\theta_{A}<\pi/$2. 

By inspection of the left panel of Figure \ref{fig:f5}, it can be seen  
for both, parallel and perpendicular directions to the LSSR major axis (red and 
black lines respectively), that high density particles (solid lines) exhibit
higher velocities than low density particles (dashed lines). In addition, 
regardless of the particle local density, lower radial velocities are obtained 
along  directions parallel to the  LSSR major axis. Therefore we recover the 
results obtained in the previous subsection regarding the general dependence 
of the radial velocity field on the LSSR. This figure also reveals a significant 
dependence of radial velocities on the local density confirming previous 
communications \citep{2006Pivato,roberto_nlsn}.
  
Galaxies and groups are expected to flow onto large clusters. 
Thus, it is interesting to study the mean radial velocity of 
haloes in the surroundings of the most massive systems. 
For the purpose of the analysis of halo dynamics we have considered
similar conditions than those applied to particles.  

For haloes we redefine the parameter characterising the local environment;
the halo local density parameter is calculated using the 100th nearest  particle
neighbour. We exclude particles within 3 virial radii in order to 
avoid the internal regions.  

We consider haloes in prolate LSSR regions and compute the averaged radial velocities
of high and low local density haloes in preferred LSSR directions. The 
results are shown in the right panel of Figure \ref{fig:f5} and consist on 
the radial velocity of low mass haloes ($7.2\times 10^{11}$M$_{\odot}$h$^{-1}<m_{halo}< 
4.5\times 10^{12}$M$_{\odot}$h$^{-1}$) in the surroundings of high mass haloes 
($1.9\times 10^{13}$M$_{\odot}$h$^{-1}<m_{halo}$). Solid (dashed) lines symbolise the infall of haloes in
high (low) local densities while red and black lines correspond to haloes along, and in the
direction perpendicular to, the major axis. 
From a general point of view we obtain radial velocities with similar 
characteristics than the velocity field traced by mass particles, namely,
low density haloes show higher velocities than haloes in dense regions.
We see an important distinction between infall of material in the 
different LSSR directions, where velocities along the major LSSR axis 
direction are larger than along the perpendicular direction.  

From the comparison of both panels of Figure \ref{fig:f5} it can be 
seen that the most prominent difference resides on the velocity 
amplitudes. Halo velocities are systematically higher than for particles.  
This difference is mainly be due to the higher mass of the central haloes, 
and it is consistent with the differences detected in Figure \ref{fig:f1}.

We have also explored the infall velocities around central haloes of different masses, and
for different LSSR shapes, and find consistent results (including higher velocities
for higher masses).

\section{Velocity dispersions and laminarity of the flow}

    \begin{figure*}
     \begin{picture}(430,250)
       \put(-40,0){\psfig{file=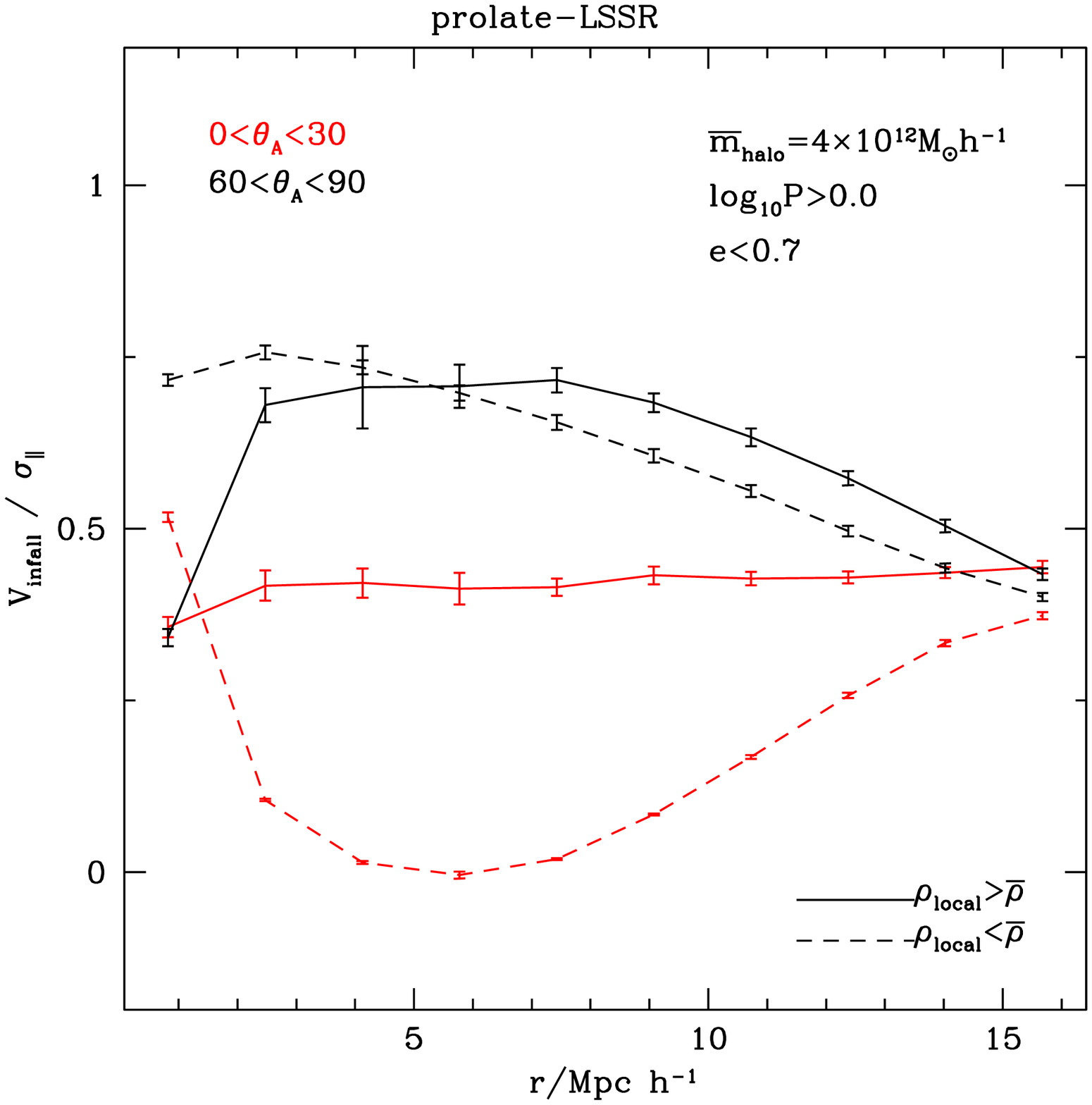,width=9.cm}}
       \put(210,0){\psfig{file=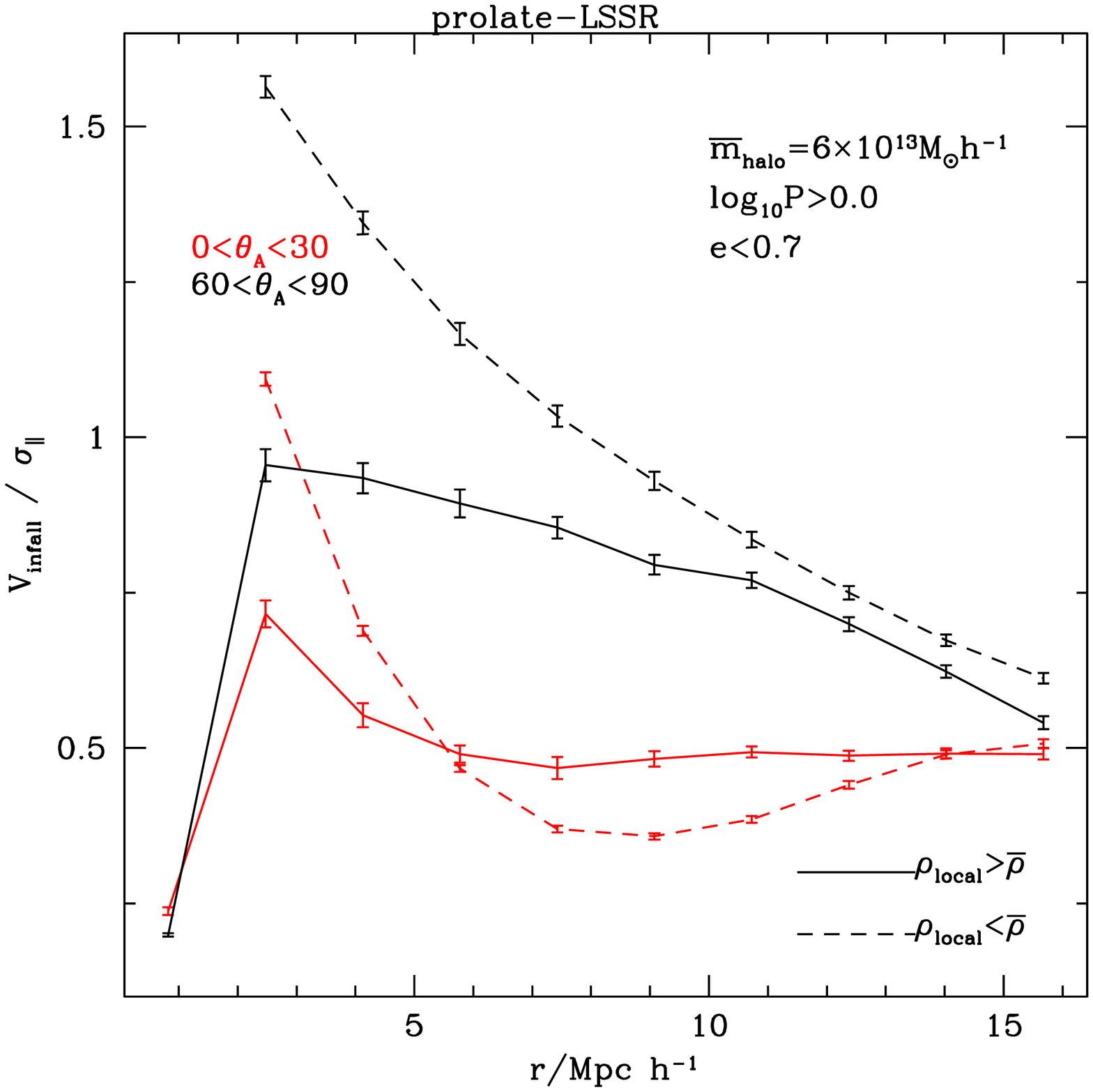,width=9.cm}}
    \end{picture}
    \caption{
    Left panel: 
    Mean laminarity parameter (the ratio between the radial velocity and the velocity 
    dispersion in the radial direction) as a function of distance for 
    particles around low mass haloes ($3.5 \times 10^{12} $M$_{\odot}$h$^{-1} < 
    m_{halo} < 4.5 \times 10^{12} $M$_{\odot}$h$^{-1}$) and prolate-LSSR 
    ($C/A < 0.7$ and $P > 0.0$). Solid lines correspond to high local 
    densities and dashed lines to low local densities, as is indicated 
    in the figure. Red lines correspond to $\theta_{A}<\pi/$6, whereas 
    black lines correspond to $\pi/$3$<\theta_{A}<\pi/$2.
    Right panel: 
    Mean laminarity parameter as a function of distance for particles 
    around high mass haloes ($m_{halo} < 1.9 \times 10^{13} $M$_{\odot}$h$^{-1}$) 
    and prolate-LSSR, lines and colors are as in the left panel. 
    }
    \label{fig:f6}
    \end{figure*}

\begin{figure*}
    \epsfxsize=18.0cm
    \centerline{\epsffile{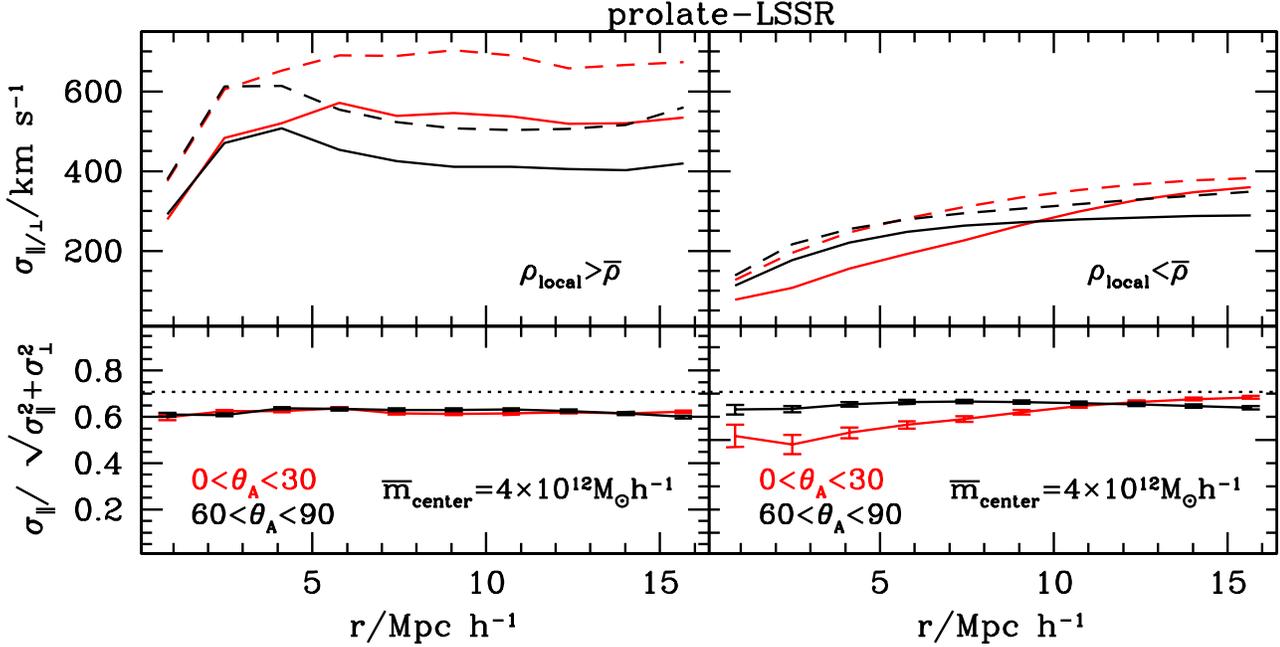}
    }
    \vspace{-7cm}
    \caption
    {
    Upper panels:
    Velocity dispersions as a function of distance for particles 
    of high and low local densities (left and right panels respectively)
    around low mass halo centres 
    ($3.5 \times 10^{12} $M$_{\odot}$h$^{-1} < m_{halo} < 4.5 \times 10^{12} $M$_{\odot}$h$^{-1}$)  
    and prolate-LSSR.
    Solid and dashed lines indicate radial and perpendicular velocity 
    dispersion respectively, and red lines correspond to $\theta_{A}<\pi/$6, 
    whereas black lines correspond to $\pi/$3$<\theta_{A}<\pi/$2.
    Low panels:
    Ratio between radial and total velocity dispersions as a function of 
    distance for particles around low mass haloes and prolate-LSSR.
    Left (right) panels correspond to high (low) local densities
    and colours indicate position angles as in the upper panels. 
    In the low panels the dotted lines show 
    $\sigma_{\parallel}/ {\sqrt{\sigma_{\parallel}^2+\sigma_{\perp}^2}} = 1/\sqrt{2}$ 
    }
    \label{fig:f13}
\end{figure*}

    \begin{figure}
     \begin{picture}(430,250)
       \put(-10,0){\psfig{file=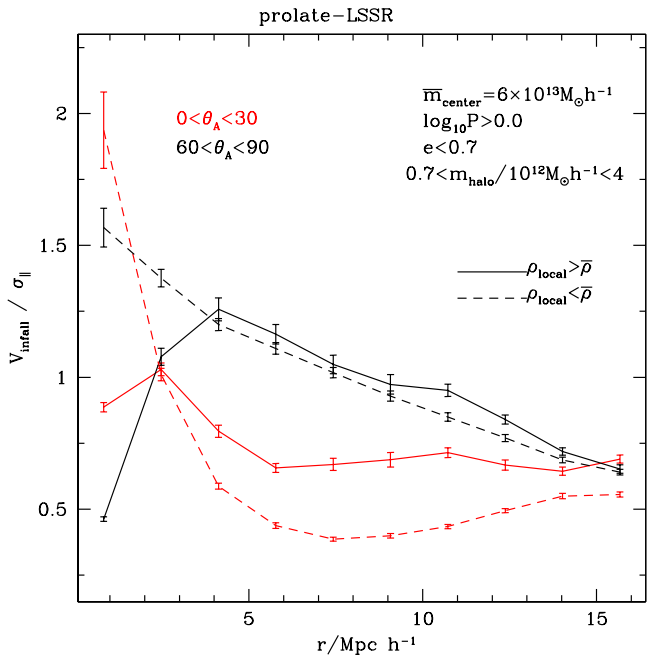,width=9.cm}}
    \end{picture}
    \caption{ 
    Mean laminarity parameter as a function of distance for low mass haloes 
    ($7.2 \times 10^{11} $M$_{\odot}$h$^{-1} < m_{halo} < 4.5 \times 10^{12} $M$_{\odot}$h$^{-1}$) 
    around high mass haloes ($1.9 \times 10^{13} $M$_{\odot}$h$^{-1} < m_{halo} $) and 
    prolate-LSSR ($C/A < 0.7$ and $P > 0.0$). Red lines indicate $\theta_{A}<\pi/$6, 
    whereas black lines indicate $\pi/$3$<\theta_{A}<\pi/$2. The two pairs of 
    solid and dashed lines correspond to high and low local densities as is 
    indicated in the figure.  
    }
    \label{fig:f9}
    \end{figure}

\begin{figure*}
    \epsfxsize=18.0cm
    \centerline{\epsffile{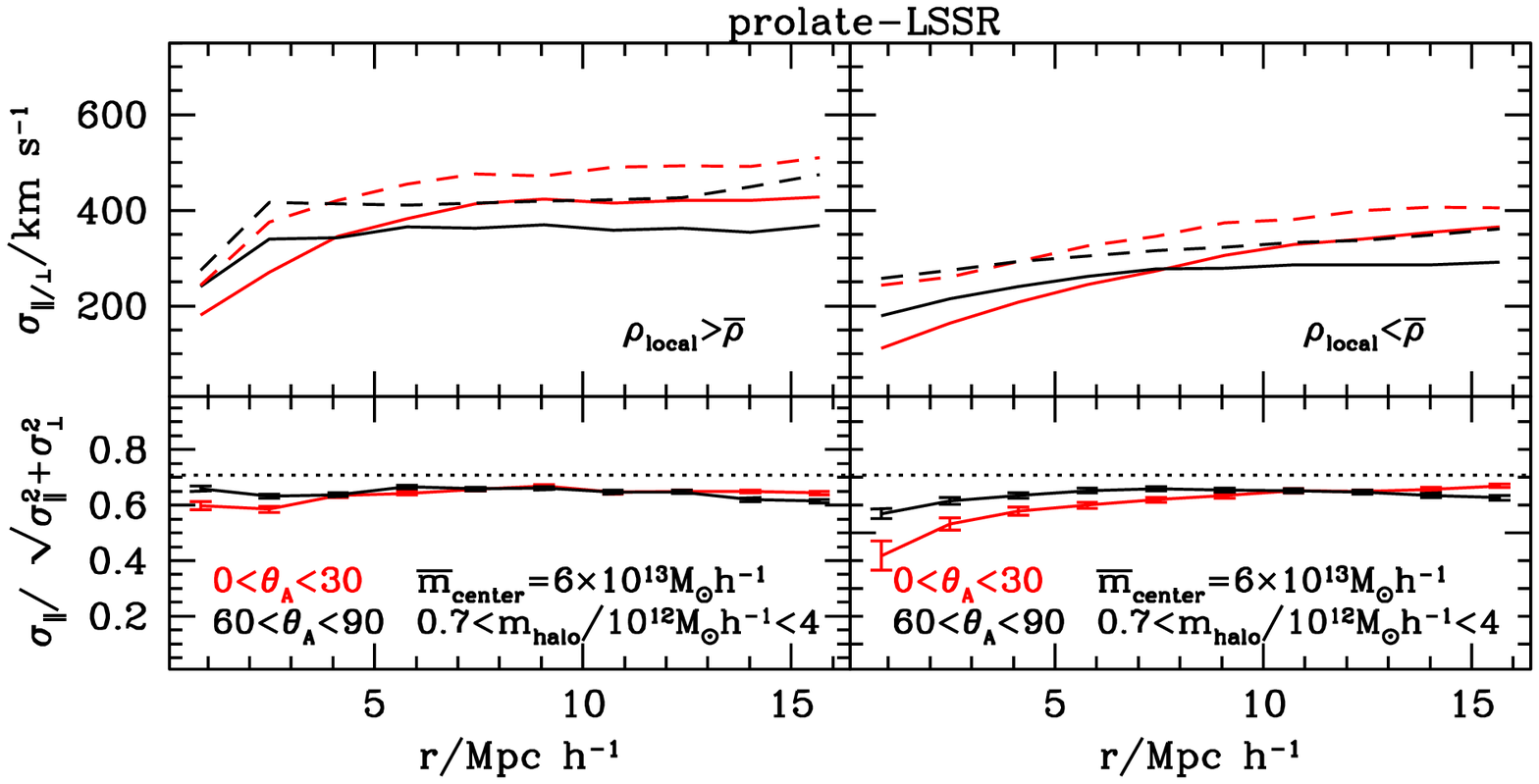}
    }
    \vspace{-7cm}
    \caption
    {
    Upper panels:
    Velocity dispersions as a function of distance for low mass haloes 
    ($0.7 \times 10^{12} $M$_{\odot}$h$^{-1} < m_{halo} < 4.5 \times 10^{12} $M$_{\odot}$h$^{-1}$)  
    of high and low local densities (left and right panels respectively)
    around high mass halo centres 
    ($ m_{halo} < 1.9 \times 10^{13} $M$_{\odot}$h$^{-1}$)  
    and prolate-LSSR.
    Solid and dashed lines indicate radial and perpendicular velocity 
    dispersion respectively, and red lines correspond to $\theta_{A}<\pi/$6, 
    whereas black lines correspond to $\pi/$3$<\theta_{A}<\pi/$2.
    Low panels:
    Ratio between radial and total velocity dispersions as a function of
    distance for low mass haloes around high mass haloes and prolate-LSSR.
    The left (right) panel corresponds to high (low) local densities
    and colours indicate position angles as in the high panels.  
    In the low panels the dotted lines show 
    $\sigma_{\parallel}/ {\sqrt{\sigma_{\parallel}^2+\sigma_{\perp}^2}} = 1/\sqrt{2}$.
    }
    \label{fig:f16}
\end{figure*}

In the previous sections we reported a significant dependence of the peculiar
velocity field on LSSR directions and local density. Our results indicate that
radial peculiar velocities deliver information on the large scale anisotropies
suggesting that the velocity dispersions could also be correlated. 

We are now interested in analysing the coherence of motions around haloes
and its relation with the infall of mass onto haloes.  To do this
analysis we define
a laminarity parameter as the ratio between radial velocity and its dispersion.  

In order to obtain a suitable laminarity parameter we compute the local
velocity dispersion for each halo using all the particles (haloes)
within spherical shells. As it was described above
we consider separately the angle subtended by the direction to a particle or halo
and the principal axes
of the LSSR.  We also take into account the
local density.

In Figures \ref{fig:f6} and \ref{fig:f9} we show the 
laminarity parameter as a function of the distance to the central 
halo, for mass particles and small haloes respectively. We only show the results 
obtained for central haloes in a prolate LSSR. In Figure 
\ref{fig:f6} we select low and high mass central haloes (left and right panels respectively), 
whereas in Figure \ref{fig:f9} we show velocities of low mass haloes around 
massive haloes.

As can be seen in the left panel of Figure \ref{fig:f6}, along perpendicular  
directions to the LSSR major axis (black lines) particles in either 
high or low local densities (solid and dashed lines) present similar 
laminarity parameters. In the case of particles along the
direction parallel to the LSSR major axis (red lines), high density particles 
(solid lines) show higher laminarity parameters than those at low local 
densities (dashed lines). We also obtain similar results using halo velocities 
instead of particles (Figure \ref{fig:f9}).
When considering particle velocities around high mass haloes 
(right panel of Figure \ref{fig:f6}), it can be noticed
that along directions perpendicular to the LSSR major axis (black lines) 
low density particles (dashed lines) show higher laminarity parameters 
than those at high local densities (solid lines). 

From these results we can infer that the flow along the different 
directions becomes more laminar (i.e. $v/\sigma > 1$) closer to the 
halo centre. This effect is significantly more pronounced for infalling 
clumps than for infalling individual particles. This laminarity is also 
larger along the plane perpendicular to the LSSR major axis indicating 
that, besides the fact that this direction shows the largest infall, it is also
the less turbulent; the material is more smoothly accreted. The 
implications of this behaviour for halo structure could reside in the 
dynamical memory of the systems of this quasi stationary process which acts
on time-scales of halo formation.

We also explore possible differences of the peculiar velocity field along
the directions parallel and perpendicular to the halo-centric direction.  In
order to analyse this we calculate the 1-D velocity dispersion of galaxies in
both the $\sigma_{\parallel}$ and, $\sigma_{\perp}$ directions, respectively, 
on the frame of the mean infall stream.  
Our results are shown in Figures \ref{fig:f13} and
\ref{fig:f16}, where we plot the velocity dispersions (top panels) and the
ratio between radial and total velocity dispersions (hereafter the relative
dispersion, bottom panels) as a function of distance, for both particles and
haloes.  Different colours indicate the position relative to the major axis
(red lines for particles/haloes along the major axis and black lines for the
perpendicular directions).  Left and right panels correspond to high and low
local densities, respectively.  Different lines in the top panels indicate
radial and perpendicular velocity dispersions (solid and dashed lines
respectively).  For comparison, the expected value for the relative dispersion
in an isotropic case is $\sigma_{\parallel}/
{\sqrt{\sigma_{\parallel}^2+\sigma_{\perp}^2}}=1/\sqrt{2}$, and is shown as
horizontal dotted lines in the bottom panels.

As can be seen from the top panels of Figures \ref{fig:f13} and \ref{fig:f16}
the dispersion of radial velocities (solid lines) is smaller than that of
velocities along the perpendicular direction (dashed lines).  This behaviour
can also be seen from bottom panels of Figures \ref{fig:f13} and \ref{fig:f16}
where the relative dispersion is lower than $1/\sqrt{2}$, regardless of the local
density, LSSR shape and distance to the central halo.  We find this behavior out to
distances as large as 15 Mpc, indicating that the radial infall dominates over
random motions. 

From the top panels of Figures \ref{fig:f13} and \ref{fig:f16} it can be
seen that the velocity dispersion for high local densities (left panels)
is higher than for low local densities (right panels), this behaviour is
stronger for particle velocities (Figure \ref{fig:f13}). With respect to the LSSR
principal axis, the velocity dispersion along the mayor axis (red lines) is
higher than in its perpendicular direction (black lines) for high local
densities at distances larger than $5$ Mpc h$^{-1}$ (left top panels of Figures
\ref{fig:f13} and \ref{fig:f16}). In general, the velocity dispersion tends to
decrease towards smaller distances from the central halo.

The lower panels of Figures \ref{fig:f13} and \ref{fig:f16}, which correspond
to low local densities, show that the relative dispersion along the major axis
decreases as the distance to the central halo decreases. At smaller distances
from the central halo ($r<10$ Mpc h$^{-1}$) the relative dispersion along the major
axis (red lines) is smaller than along the perpendicular direction (black
lines), whereas at larger distances ($r<13$ Mpc h$^{-1}$) we find the opposite
behaviour. We note that the dependence of the relative dispersion on the
direction with respect to the
LSSR major axis tends to be smaller as the distance increases
(as expected due to the tendency towards homogeneity of the LSSR).  

The relative dispersion for high local densities (left bottom panels of Figures
\ref{fig:f13} and \ref{fig:f16}) shows similar behaviours for the different LSSR
directions and, as it can be seen, there is little dependence on the distance
to the central halo with a roughly constant value $\sigma_{\parallel}/
{\sqrt{\sigma_{\parallel}^2+\sigma_{\perp}^2}} = 0.6$.  We do not detect a
strong dependence of the relative dispersion on local density and LSSR.
A more detailed analysis shows a slight tendency of the
relative dispersion to be smaller along the radial than the perpendicular
direction, for both, particles and haloes, as is seen in the right bottom
panels of Figures \ref{fig:f13} and \ref{fig:f16}. 

\section{Infall and outflow}

    \begin{figure*}
    \epsfxsize=14cm
    \centerline{\epsffile{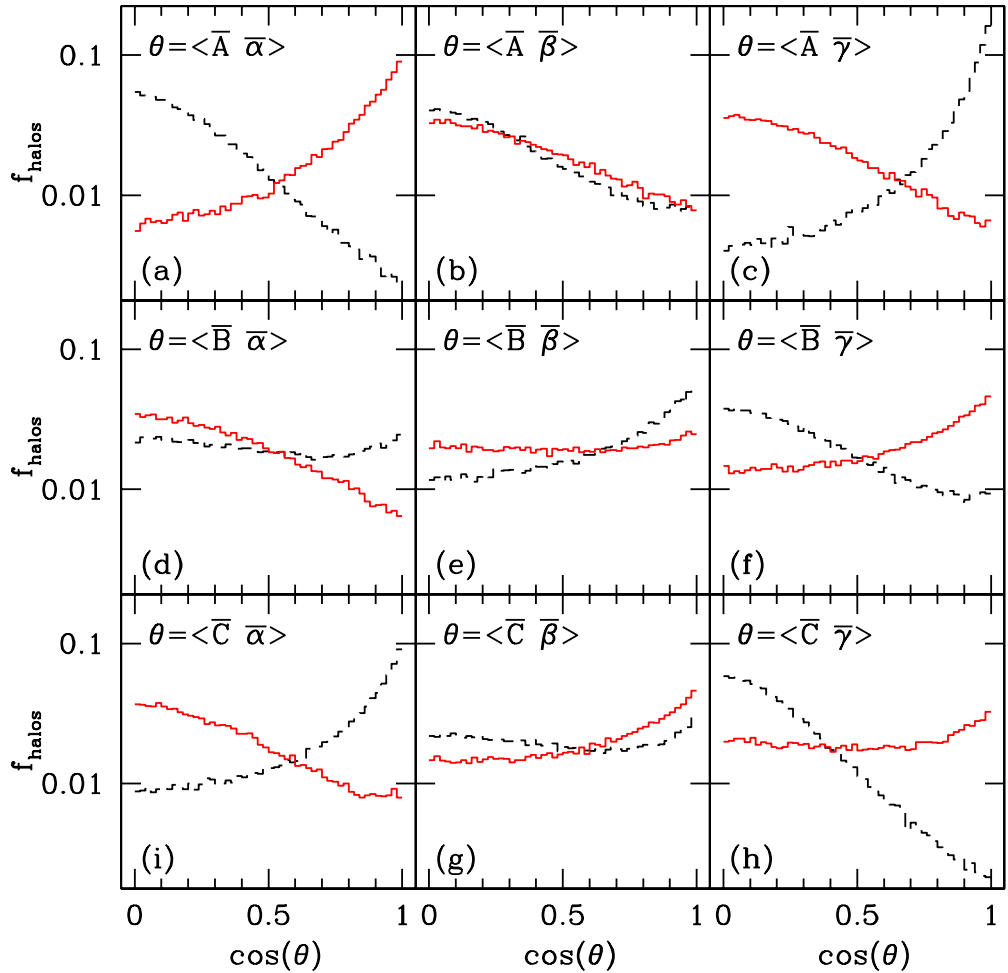}
    }
    \caption{Histograms showing the vectorial products between 
     principal axes derived from the infall 
     velocity tensor and the LSSR for particles (dashed black lines). 
     Each panel represents the angle between different pairs of vectors as is 
     indicated in the corresponding panel in the figure. 
     A, B, and C represent  the
     major, intermediate and minor axes of the LSSR, respectively.
     $\alpha$, $\beta$ and $\gamma$ represent  the 
     major, intermediate and minor axes of the infall velocity ellipsoid, respectively. 
     The corresponding histograms derived from the outflow velocity tensor 
      and the LSSR are shown as solid red lines. 
    }
    \label{fig:f18}
    \end{figure*}

In this section we provide further statistical properties of matter infall to, 
and outflow from haloes. For this aim we have considered
separately particles with negative (infalling) and positive (outflowing) radial
velocity components.  Therefore, these two groups of particles consist in local
diverging and converging flows onto the halos.

Velocity moments for each group of converging or diverging particles are
obtained in a similar way as the one used to obtain the shape tensor (Equation
\ref{eq:it}), using the components of the peculiar velocity vector of each
particle relative to the centre of mass velocity of the surrounding LSSR (within
a sphere of 15 Mpc h$^{-1}$ from the halo centres) instead of the position vector. 

Then, for each group of particles we obtain a triaxial ellipsoid that
approximates the velocity distribution. The corresponding major, intermediate
and minor principal velocity axes are  referred to as $\alpha$, $\beta$ and $\gamma$.  We
analyse the orientation of these two velocity ellipsoids (for infalling and
outfalling particles), with respect to the principal axes of the surrounding
structure. As in previous sections, we refer to these spatial axes as $A$, $B$
and $C$, for the major, intermediate and minor axes, respectively.  We compute
the angles subtended by the principal axes of shape and velocity ellipsoids and
summarise our results in Figure \ref{fig:f18}.  This figure shows the
distributions of the cosine of the angles between LSSR and velocity axes, for
converging (black lines) and diverging particles (red lines). Each panel in
this figure corresponds to a different pair of axes, the three upper panels
correspond to angles between the LSSR major axis and the velocity major,
intermediate and minor axes (panels a, b and c, respectively). Similarly, the
three middle panels correspond to the LSSR intermediate axis and the three lower
panels for the LSSR minor axis, as indicated in the figure. 

As can be seen in the left upper panel of this figure, the major LSSR and
velocity axis tend to be perpendicular for converging particles (black lines in
panel (a)), and parallel for diverging particles (red lines in panel (a)). The
right upper panel of this figure shows that the major LSSR axis tends to be
aligned with the minor velocity axis for converging particles (black lines in
panel (c)) whereas for diverging particles they tend to be perpendicular (red
lines in panel(c)).  Regarding the angles between the LSSR major axis and the
velocity intermediate axis there are no significant differences between
converging and diverging particles (black and red lines in panel (b)).  By
inspection of the lower panels of this figure, which correspond to the angles
between the LSSR minor axis and the velocity ellipsoid, it can be seen,
consistent with the results in the upper panels, that the minor LSSR axis and
the major velocity axis are aligned for converging particles and perpendicular
for diverging particles (black and red lines in panel (i)). 

The main results shown in Figure \ref{fig:f18} indicate that LSSR major axes
tend to be parallel to major (minor) velocity axes for diverging (converging)
particles. The alignment signals for the converging flow is consistent with the
results derived in the previous sections,  namely a preference for larger
amplitudes of the infalling velocity field along directions perpendicular to
the principal axes of the surrounding structure. 

Regarding the outflowing velocity particles we notice the opposite behaviour,
the largest amplitude occurs along the LSSR major axis direction. Thus, in part,
the reduced mean infall along the LSSR major axis could be caused by the
preference of outflowing particles along this direction. We can interpret this
trend as generated by particles expelled from the virialised structures, which
could occur more easily along the major axis where the gravitational potential
shows a more gentle decrease than along the minor axis direction.  Thus, a
possible implication of our results would be the presence of galaxies that
passed close to the halo centres and are outflowing along the surrounding
filamentary and planar structures.

\section{CONCLUSIONS}

Using a numerical simulation we have performed a detailed analysis of the
peculiar velocity field around dark matter haloes. We have explored the
relation between the flows of mass towards haloes and the shapes of the central
haloes, their large-scale surrounding region (LSSR) shape, and their orientation. 

We found a significant anisotropy in the velocity field which correlates with
the surrounding LSSR principal axes. The amplitude of the velocity field of
infalling particles onto dark matter haloes is maximum along the direction of
underdense surrounding regions (LSSR minor axis), whereas along the direction of
dense LSSR regions, it exhibits the smallest velocities.  With respect to the
shape ellipsoid of the central haloes, we found that the amplitude of the
infall velocity field along the halo minor axis is larger than that along the
major axis. We find consistent results for general triaxial haloes, and for
both prolate and oblate systems. Therefore, there is a clear preference for
matter to infall towards haloes along regions of low ``global" density.  

When looking at the dependence with local density, either measured in cubic
cells in the simulation volume for individual particles, or using the 100th
closest particle neighbour outside $3$ virial radii for tracer haloes, we find
that objects in local high density regions infall faster.
 
By defining a suitable laminarity parameter we find that the velocity field is
less turbulent along the direction towards underdense surrounding regions, with
a ratio between mean flow velocity and velocity dispersion of order unity and
nearly constant up to scales of 15 Mpc h$^{-1}$.

The flow of particles moving away from haloes are found to be highly
anisotropic. The major axis of the outflowing velocity field 
is aligned with the surrounding LSSR major axis.

\section*{Acknowledgments}
This work has been partially supported by Consejo de Investigaciones
Cient\'{\i}ficas y T\'ecnicas de la Rep\'ublica Argentina (CONICET), the
Secretar\'{\i}a de Ciencia y T\'ecnica de la Universidad Nacional de C\'ordoba
(SeCyT).

\end{document}